\documentclass{vgtc}                     % final (journal style)
\onlineid{1022}

\vgtccategory{Research}

\vgtcpapertype{algorithm/technique}

\title{Visualizing Uncertainty in Non-linear Projections with Ensembles}
\author{Kai Nylund\thanks{e-mail: nylund.k@northeastern.edu}\\ %
        \scriptsize Northeastern University %
\and Michael Correll\thanks{e-mail: m.correll@northeastern.edu}\\ %
     \scriptsize Northeastern University %
\and Lace Padilla\thanks{e-mail: l.padilla@northeastern.edu}\\ %
     \parbox{1.4in}{\scriptsize \centering Northeastern University}}

\abstract{
    Widely used non-linear dimensionality reduction (NLDR) methods such as UMAP and t-SNE are stochastic---repeated runs on the same data can produce different low-dimensional projections. In this paper, we explore two problems related to projection variability: on some datasets clusters, structure, and outliers may change run-to-run, and on others projections can be extremely stable when overfitting noise. To address the first problem, we propose visualizing the median of multiple NLDR outputs rather than relying on individual projections. To address the second, we perturb input data before creating consensus embeddings. We find that taking the median of multiple projections performs comparably to individual runs on multiple quality metrics, while increasing perturbation emphasizes global over local structure. We show through a set of exploratory visualizations that even relatively simple ensemble presentations can be used to better communicate the reliability of projection patterns.
}


\teaser{
  \centering
  \includegraphics[width=\linewidth, alt={A flowchart depicting ensembling with multiple projections. A small scatterplot is on the left with the title "high-dimensional data". There is a solid arrow from that scatterplot to multiple overlapping 2D projections of a real dataset with the title "generate multiple projections". There is also a dotted arrow between the two with the title "perturb with noise or sampling". Finally, there is an arrow from the multiple projections to examples of our five ensemble visualizaiton types (median projection, small multiples, confidence ellipses, movement lines, and KDE + binned points) with the title "visualize as an ensemble".}]{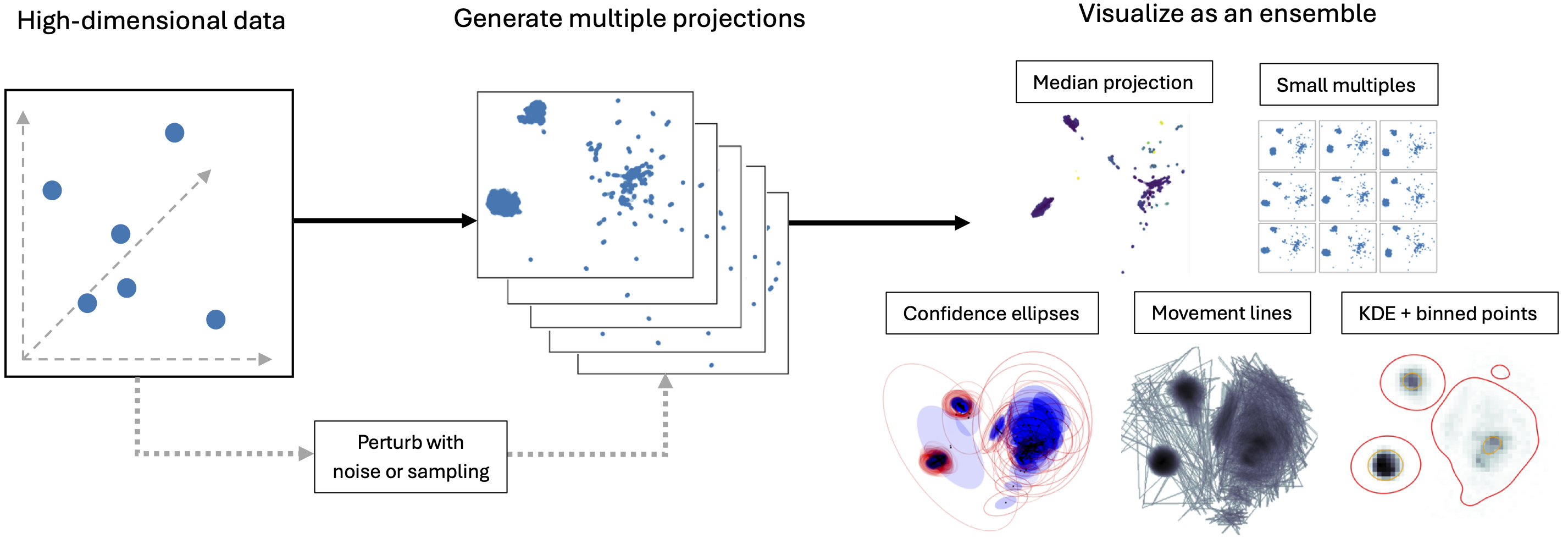}
  \caption{\textbf{Visualizing ensembles of non-linear projections reveals uncertainty lost in individual outputs.} We generate several projections for the same dataset to identify how they can change run-to-run. To avoid stable but spurious projections, we optionally perturb input data by adding noise or subsampling. Finally, we explore five ensemble visualization methods to communicate variability.}
  \label{fig:teaser}
}

\graphicspath{{figs/}{figures/}{pictures/}{images/}{./}} % where to search for the images

\usepackage{mathptmx}                  % use matching math font

\usepackage[table,xcdraw]{xcolor}
\usepackage{tabularx}
\usepackage{amsmath}
\usepackage{amssymb}
\usepackage{comment}
\usepackage{enumitem}
\usepackage{hyperref}
\usepackage[subtle]{savetrees}

\begin{document}

%%%%%%%%%%%%%%%%%%%%%%%%%%%%%%%%%%%%%%%%%%%%%%%%%%%%%%%%%%%%%%%%
%%%%%%%%%%%%%%%%%%%%%% START OF THE PAPER %%%%%%%%%%%%%%%%%%%%%%
%%%%%%%%%%%%%%%%%%%%%%%%%%%%%%%%%%%%%%%%%%%%%%%%%%%%%%%%%%%%%%%%

\maketitle

\section{Introduction}
\label{sec:intro}

Non-linear dimensionality reduction (NLDR) techniques like t-SNE \cite{van2008visualizing} and UMAP \cite{mcinnes2018umap} are widely used to visualize and interpret high dimensional data. NLDR methods are also often stochastic---repeated runs on the same data can produce different low-dimensional projections. In this work, we highlight two issues related to NLDR stability: (1) on some datasets structure, clusters, and outliers may change run-to-run, and (2) on others, projections are extremely stable when overfitting noise. Combined with other well-documented drawbacks such as hyperparameter tuning \cite{wattenberg2016use} and the inclusion of spurious patterns \cite{wattenberg2016use, chari_specious_2023}, we argue that failing to communicate this stochasticity can lead to \textbf{overinterpretation errors} including confirmation bias and multiple comparisons problems. 

Overinterpreting visual patterns in NLDR can lead analysts to draw confident but unsupported conclusions about the structure of high-dimensional data. For example, a study aiming to sequence genomes from 1 million volunteers across the United states \cite{all2024genomic} drew criticism for using UMAP to project whole genome sequence data colored by self-reported race (see figure \href{https://www.nature.com/articles/s41586-023-06957-x/figures/2}{here}). The figure, which contains what appear to be clearly defined clusters based on whether participants self-reported as Black, Asian, white, or another racial group, was criticized for falsely implying that race is biologically determined \cite{kozlovAllofUs}, simplifying the complexity of both genetic ancestry and the construction of race~\cite{omi2020racial}. The seemingly well-defined clusters are an artifact of UMAP but uncertainty about whether the scatterplot points accurately represent the underlying data is not communicated to readers.

Prior work commonly frames NLDR instability as \textit{unreliability} and tries to reduce \cite{wang2021understanding, jeon2025umato}, or quantify \cite{liu2025assessing, jung2025ghostumap2} projection variability. We argue, however, that instability is meaningful uncertainty that should be made visible to analysts. Yet, in practice variation is not communicated by existing NLDR visualization and evaluation methods (mainly scatterplots and quality metrics). As a result, analysts may reasonably overinterpret patterns that are artifacts of a particular projection rather than stable features of the data. In contrast, communicating instability may help analysts avoid overinterpreting clusters, distances, and structure. 

In this work, we explore this challenge in NLDR uncertainty communication by visualizing projection variation with ensembles of outputs. Adapting uncertainty quantification and visualization techniques from other domains~\cite{wang2018visualization}, we visualize the results of an \textit{ensemble} of NLDR runs under perturbed initial conditions. The resulting ensemble visualizations communicate the reliability and uncertainty of positions of points, clusters, and higher-level structure. We make three contributions:

\begin{enumerate}[noitemsep,nolistsep]
    \item Applications of methods for ensembling NLDR outputs and visualizing NLDR ensembles.
    \item Demonstration of how ensembling NLDR outputs can maintain or improve projection quality metrics, and may help reduce confirmation bias and spurious patterns.
    \item 1M+ projections of well-known datasets publicly available for future study of NLDR stability and ensembling.
\end{enumerate}

\begin{figure}[t!]
    \centering
    \includegraphics[width=\linewidth, alt={Depictions of the two main problems we explore in the paper. A title in the top left reads "Problem 1: projection variance can cause multiple comparisons problems", with a 3d scatterplot of original data (two concentric spheres) and nine example projections to the right of that colored to show different numbers of clusters. A title on the top right reads "Solution: visualize projection ensembles", with the median of multiple projections shown as a scatterplot below. Text under the median reads "Taking the medin of multiple projections reveals two original clusters". A title on the bottom right reads "Problem 2: projections can be extremely stable when overfitting noise", with another 3d scatterplot of original data (a gaussian blob) and overlaid scatterplots with a single projection and the median projection showing little variation between the two and contains areas of high and low density. Text reads "Median ensemble shows spurious patterns due to stability". Finally, in the bottom right is the title "Solution 2: perturb data before ensembling", showing the median ensemble + 5\% noise, which more closely resembles the underlying gaussian distribution.}]{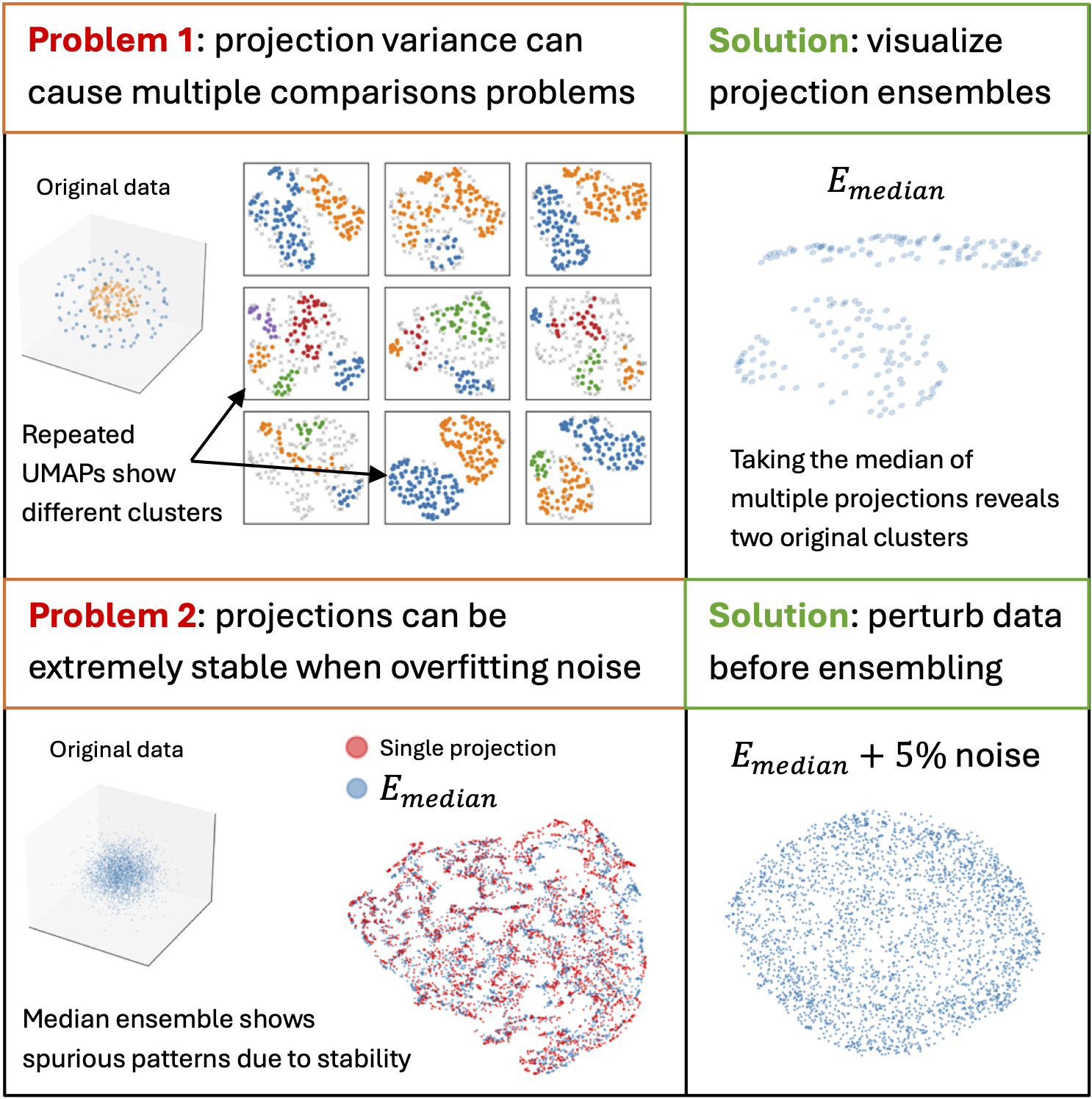}
    \caption{We highlight two key problems related to NLDR stochasticity: changing structure between runs (top left), and stability when overfitting noise (bottom left). We explore addressing these issues by taking the median of multiple projections (top right), and perturbing data points before generating projection ensembles (bottom right) respectively. Clusters in the top left are colored with HDBSCAN on the 2D projection to show perceptual variability.}
    \label{fig:problems_summary}
    \vspace{-2mm}
\end{figure}
\section{Related Work}
\label{sec:related_work}

\subsection{Ensemble visualization}
Rather than presenting the outputs of a single model, one can instead visualize an \textit{ensemble} of models' outputs. Ensemble techniques have shown promise for improving the reliability and robustness of models in a wide range of domains, particularly in weather and climate modeling~\cite{gneiting2005weather}. These ensembles, however, result in an increase in the complexity of the data to be visualized. Researchers have therefore developed many approaches for visualizing ensemble data (for a review, see~\cite{wang2018visualization}). Hurricane forecast tracks, for instance, could aggregate all model information into a single ``cone of uncertainty'', disaggregate these data into individual predicted paths~\cite{ruginski2016non}, or compromise between these two extremes by including information about the overall distribution of the ensemble augmented by representative~\cite{liu_uncertainty_2017} or outlying~\cite{whitaker_contour_2013} ensemble predictions. In this work we explore a set of ensemble visualizations in this design space, applied to the domain of NLDR robustness.

\subsection{NLDR unreliability}
Dating back to the earliest uses of principal component analysis (PCA), analysts have a history of misinterpreting dimensionality-reduction techniques~\cite{gould2012mismeasure}. Modern non-linear methods present more opportunities for misunderstanding because axes often lack semantic meaning and mappings can be difficult to interpret~\cite{gleicher2013explainers}. Recent work has drawn attention to these limitations and cautioned against using NLDR methods to draw conclusions based on point distances and label distributions \cite{chari_specious_2023}. 

Although researchers have made numerous advances aimed at addressing these problems, many have caveats and drawbacks. For example, new NLDR methods often still suffer from issues of uninterpretability and hyperparameter sensitivity \cite{xiang2021comparison, jeon2025umato}. Quality metrics can help quantify when projections are faithful or preserve useful properties, but have also been shown to be easy to manipulate, and sometimes difficult to interpret without additional information \cite{machado2025necessary}. Furthermore, existing distortion visualization methods can help show \textit{when} and \textit{where} projections may be unfaithful or unstable, but not \textit{how} they may vary across runs (see ~\cite{jeon2025unveiling} for a review).

\subsection{Visualizing projection uncertainty}

A growing body of literature has examined how to visualize uncertainty and variation in dimensionality reduction techniques, each with strengths and weaknesses. \textit{Layerflow} visualizes differences between projections by aligning them in a row and then connecting points in clusters with sankey flows~\cite{sevastjanova_layerflow_2025}. However, the resulting flows depend on the ordering of charts and do not scale to large numbers of projections given a limited page width. Jung et al. propose a new method to assess the stability of UMAP projections by identifying individual points that are unstable under perturbation \cite{jung2025ghostumap2}. Similarly, Liu et al., perturb individual points in input data across runs to measure changes in embeddings, assigning ``perturbation'' and ``singularity'' scores to quantify local and global distortions \cite{liu2025assessing}. 

Prior work has also proposed methods to specifically visualize the \textit{stability} of non-linear projections by, for example, overlaying Voronoi diagrams \cite{reinbold2020visualizing}, highlighting robust structures \cite{jung2023projection}, and plotting hypothetical outcomes under input uncertainty \cite{zabel2025visualizing}. Our work is complementary to these recent advances. Beyond pointwise and regional stability, we apply new ensembling visualization methods to show variation across whole projections and reduce overinterpretation.

Other works aim to create dimensionality reduction techniques that can handle uncertain inputs \cite{laakom_non-linear_2022}, including uncertainty-aware versions of PCA \cite{gortler_uncertainty-aware_2020}, t-SNE \cite{ma_uncertainty-aware_2024}, and MDS \cite{hagele_uncertainty-aware_2022}. Although these can effectively represent quantified \textit{aleatoric} uncertainty (i.e., inherent irreducible randomness) in inputs, they do not account for \textit{epistemic} uncertainty (arising from e.g., limited knowledge, modeling, and visualization)~\cite{spiegelhalter2017risk}. The latter form of uncertainty is the focus of this work.
\begin{figure}[t!]
    \centering
    \includegraphics[width=\linewidth, alt={A horizontal box plot titled "UMAP projection variance across 31 datasets" with dataset on the y-axis and average procrustes distance from each projection a reference on the x-axis from 0 to 1. The datasets 11 100d spheres and 5k word vecs. from Olmo2 have the highest procrustes distances with means around 0.8; while several datasets such as 3 Gaussian blobs, Fashion MNIST, and roll w/ hole 3k have mean procrustes distances close to 0.}]{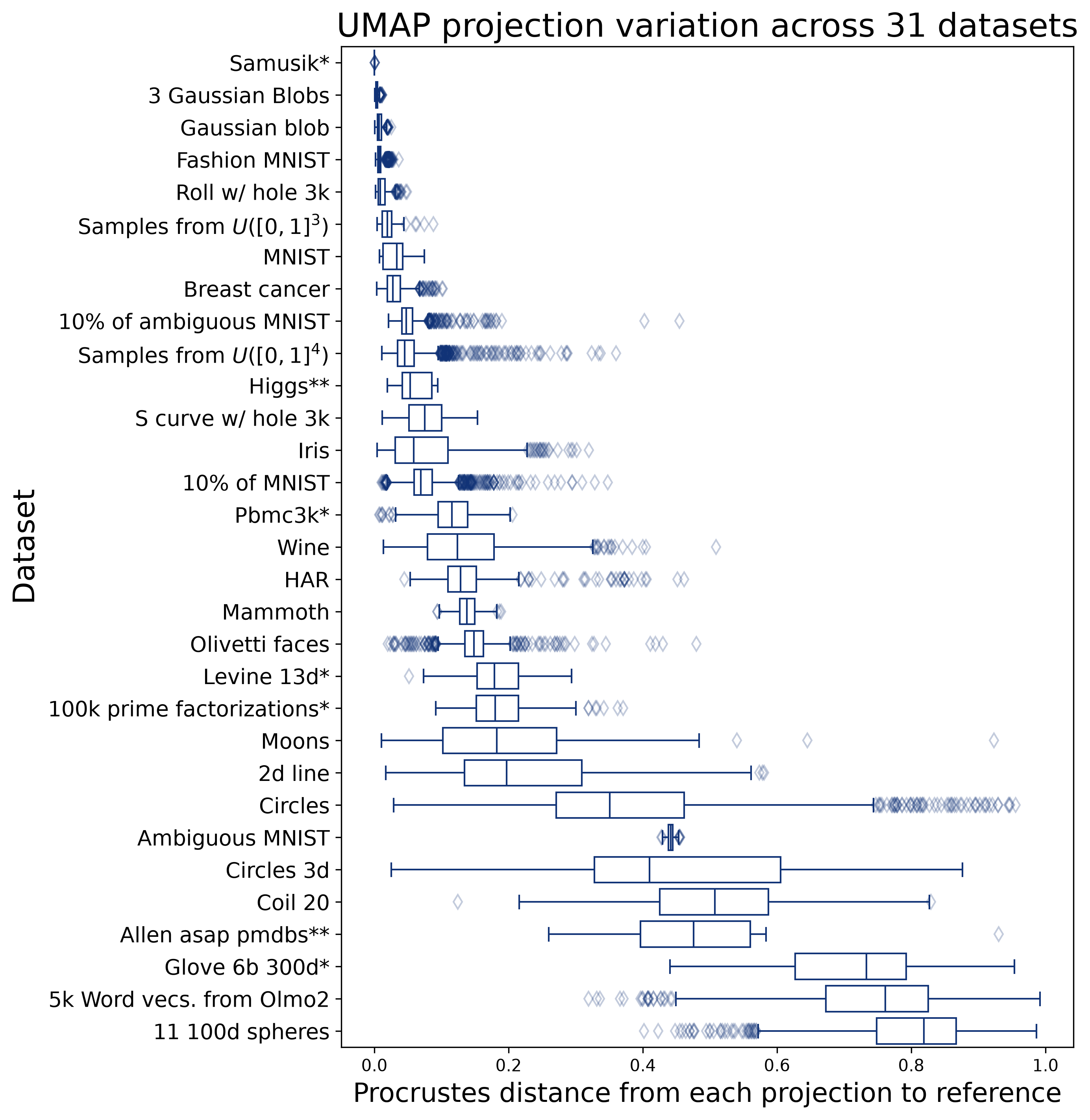}
    \caption{\textbf{Projection stability can vary widely across different datasets.} Procrustes distances between multiple UMAP runs and a reference projection across 31 datasets. Based on dataset size: no star indicates 1000 projections, *=100 projections, and **=10 projections.}
    \label{fig:procrustes_distances}
    \vspace{-1mm}
\end{figure}

\section{Projection Stability and Its Interpretive Risks}
In this section, we analyze the stability of UMAP projections across 31 datasets (full list in supplemental material), and provide examples of two central issues illustrated in~\autoref{fig:problems_summary}: (a) clusters and structure changing run-to-run, and (b) extreme stability of projections which overfit noise. For this initial exploration, we limit our experiments to UMAP with default hyperparameters (n$\_$neighbors=15, min$\_$dist=0.1). Our goal is not to conduct a comprehensive analysis of NLDR methods' stability, but rather to demonstrate the existing range of projection variability and resulting interpretation risks. For each dataset, we generate 1000 projections if there are fewer than $100,000$ data points, 100 if there are fewer than 2 million, and 10 otherwise (indicated with stars in \autoref{fig:problems_summary}). We measure the normed Procrustes distance for each run using the first projection as a reference. We provide more information about each dataset in supplemental material. The code and projections generated in this work are available at \url{https://github.com/KaiNylund/proj-ensembles}.

The average Procrustes distance and distance variability vary greatly across datasets (\autoref{fig:procrustes_distances}). 
%For UMAP, the stability of embeddings correlates with how well the datasets fit a manifold, with those that can be fit with high accuracy (e.g., blobs, swiss roll, and MNIST) settling on a minima. While others (e.g., circles, prime factors, and 11D spheres) do not converge and show greater variation.
\autoref{fig:problems_summary} Problem 1 is an example of how this variation on the circles-3D dataset results in a different number of perceptual clusters and outliers across runs. Our proposed solution is to take the median of multiple projections (\autoref{fig:problems_summary} Problem 2 Solution), revealing clusters present in the original data without selectively choosing a favorable run. On the other end of variation,~\autoref{fig:problems_summary} Problem 2 shows how UMAP projections of a 3D Gaussian distribution are highly stable--with the median of 1000 Procrustes-aligned runs almost identical to a single projection--despite overfitting noise in the data and displaying spurious areas of higher and lower density. As a potential solution, we explore ensembling with perturbation (\autoref{fig:problems_summary} Problem 2 Solution) in Section \ref{sec:ensemble_preprocessing}.

Together, these situations arising from NLDR projection stochasticity (or lack thereof) can lead to overinterpretation issues. In the first scenario, various analysts seeking to answer the question ``how many clusters are in the data?'' may come to different conclusions even when they use the exact same hyperparameters and initialization to generate standalone projections. In the second, if an analyst is aware of UMAP variability and generates several projections of the underlying dataset, they may come to the false conclusion that lack of variation means the displayed structure is present in the original high-dimensional data.

\begin{comment}
For UMAP at least, the stability of embeddings appears to correlate with how well the datasets can be embedded on a manifold, with those that can be fit with high accuracy settling on a minima while those that show greater variation. 

Interestingly, ambigious-MNIST, whose uncertainty stems more from label uncertainty than ability to fit on a manifold, shows high average variation but low variance in difference -- i.e., all plots are different but that difference is predictable.

\begin{enumerate}
    \item  NLDR projections can vary widely from run-to-run (show examples of this leading to different interpretations of data, multiple comparisons problems (structure is shown in cherry-picked projection but not others)).
    \begin{itemize}
        \item Examples
        \item Quantifying variation in projections
    \end{itemize}
    \item NLDR projections can be very certain about spurious patterns – reproducing the same wrong result in all projections due to overfitting. 
    \begin{itemize}
        \item Quantifying the lack of variation (can use previous example). “Our earlier result also points to another problem: several of the datasets have exceptionally low variation across runs – but that does not mean they are accurate representations.
    \end{itemize}
\end{enumerate}
\end{comment}
\begin{figure}[t!]
    \centering
    \includegraphics[width=\linewidth, alt={A heatmap of average metric rankings for the median ensemble with different levels of noise and dropout compared to individual runs. Ensemble method is on the y-axis including individual projections, E_median, noise percentages from 1-100, and dropout percentages from 1-99.9. 18 projection metrics implemented in the ZADU library are on the y-axis. The z-axis shows ranking from best (lightest) to worst (darkest) using a yellow-purple color scale. The title indicates columns on the left correspond to local metrics, those in the middle are cluster-level, and those on the right are global. Individual projections appear to do the best on individual runs, followed by E_median and low-levels of noise. Higher levels of dropout appear to do better on some cluster-level and global evaluation metrics.}]{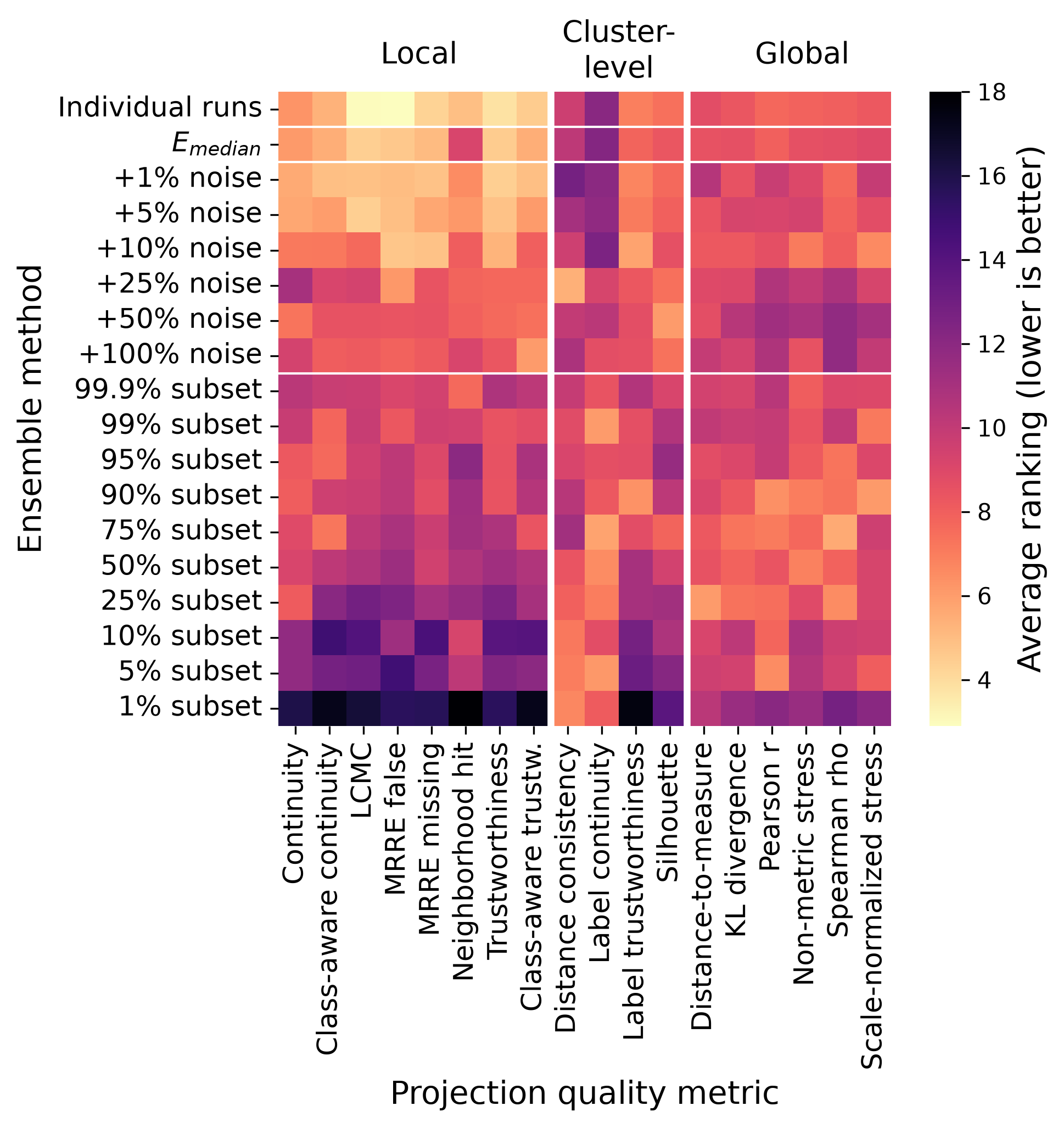}
    \caption{\textbf{Ensembles perform comparably to individual projections across projection quality metrics. Increasing perturbation prioritizes global over local performance.} We compare ensembling methods across 31 datasets with varying levels of added noise and subsampling. We report each method's average ranking compared to the others on local, cluster-level, and global projection quality metrics (lighter is better).}
    \label{fig:metric_rankings}
    \vspace{-1mm}
\end{figure}

\begin{figure*}[t!]
    \centering
    \includegraphics[width=\textwidth, alt={An overview of the five ensemble visualization methods we explore with tradeoffs. On the left is median projection with the text "avoids cherry-picking and shows how much but not how points move". Next (to the right) is small multiples with the annotation "allows for global but not local comparison between runs". Next is confidence ellipses with the text "shows point, cluster, and global uncertainty, but not fine-grained movement". Next is movement lines, with the text "shows outlier movements and cluster certainty, but not individual points". Finally, on the right is KDE + binned points with the text "Highlights consistent clustering and density but not points or movement".}]{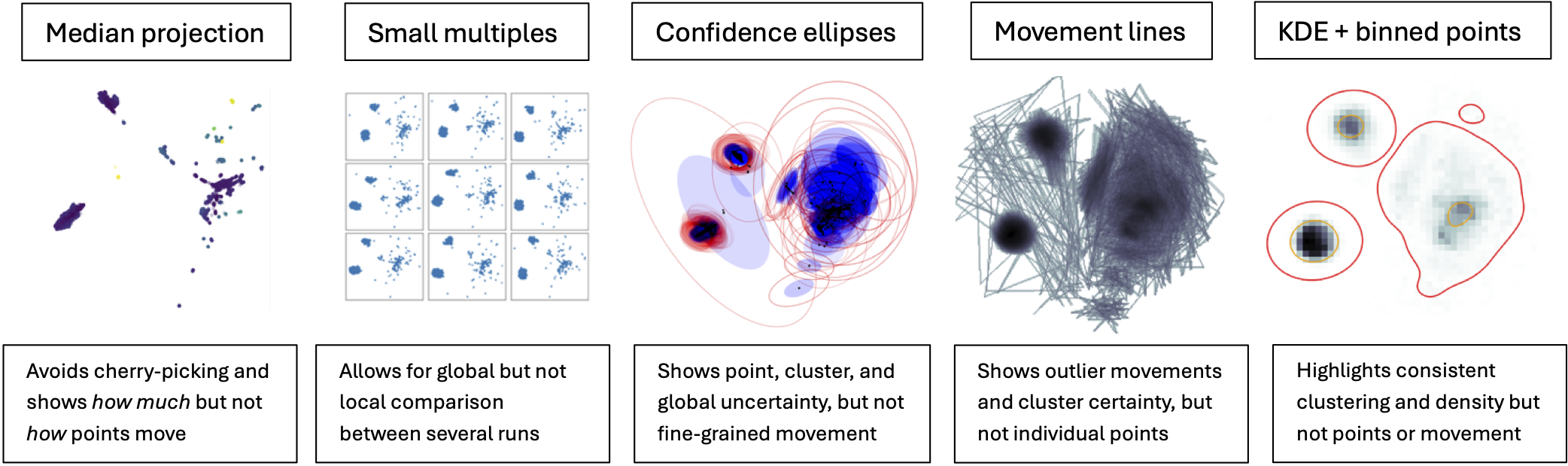}
    \caption{\textbf{Ensemble visualization methods reveal distinct types of variability across projections.} Examples of each of the five ensemble visualizations we explore on the HAR dataset \cite{Anguita2013APD}. The bottom row describes tradeoffs for each method.}
    \label{fig:ensemble_vis_types}
    \vspace{-2mm}
\end{figure*}

\section{Ensembling and Visualizing NLDR Projections}
\label{sec:ensemble_methods}

We explore methods to better communicate projection stochasticity through the use and visualization of ensembles. We first explore simple methods to create consensus embeddings and perturb inputs to avoid overfitting, then evaluate whether ensembles offer improvements over individual projections on several quality metrics. Finally, we explore five different methods to visualize projection ensembles.

\subsection{Ensembling multiple projections of the same dataset}
\label{sec:ensemble_preprocessing}

In this section, we explore creating consensus embeddings from multiple projections of the same dataset. Given $k$ projections with varying random seeds for initialization, the most straightforward way to create an ensemble is to average over all samples. To create this mean ensemble ($E_{\text{mean}}$), we first align each projection to a reference based on Procrustes distance, then compute the pointwise mean over all runs. We similarly construct ensembles by taking the pointwise median of all runs ($E_{\text{median}}$), which has the benefit of guaranteeing that each ensemble point is included in at least one projection rather than a linear combination of positions from several projections. For example, in the case of bimodal outputs from a NLDR method, $E_{\text{mean}}$ would yield points that interpolate between the two variations while $E_{\text{median}}$ would not. We also calculate the trimean ($E_{\text{trimean}}$) and geometric median ($E_{\text{geomean}}$) ensembles. Because we found few meaningful differences between the four strategies, we focused our following evaluation on $E_\text{median}$ as it is the simplest while preserving (non-interpolated) point positions directly output by NLDR methods.

To address the issue of overly stable spurious projections, we explore perturbing projections before ensembling -- a common strategy in literature on both ensembling \cite{wang2018visualization} and projection quality \cite{liu2025assessing, jung2025ghostumap2}. We use two perturbation strategies: adding noise to and subsampling input data. Given $n$ $d$-dimensional points and a projection map $f: \mathbb{R}^{n, d} \rightarrow \mathbb{R}^{n, 2}$ we add $u$\% normally-distributed noise to input data with standard deviation $\sigma$ by computing $f(X + \frac{u}{100}*\mathcal{N}(0, \sigma^2))$. We then ensemble by taking the median of outputs. As an alternative to perturbing with noise, we subsample $v$\% of data points before projecting and take the pointwise median over only samples containing those points. Assuming $k(\frac{v}{100}) > 1$, we chunk samples from the full dataset to ensure that each point is included in at least $\lfloor k(\frac{v}{100}) \rfloor$ projections. This subsampling results in fewer references for each point, but has the benefit of reducing computational cost, particularly at low subsampling ratios, and avoiding altering potentially important dataset structure (e.g., by adding noise to binary vectors).

\subsection{Evaluating projection ensembles}
\label{sec:ensemble_evaluation}

Although our goal is to reduce overinterpretation rather than create the best consensus embedding, we found that simple ensembling methods can improve over individual projections on several quality metrics. We evaluated $E_\text{median}$ ensembles on projection quality metrics implemented in the Python ZADU \cite{jeon_zadu_2023} library using a range of different added noise $u \in [0.1, 1, 5, 10, 25, 50, 100]$ and subsampling $v \in [99.9, 99, 95, 90, 75, 50, 25, 10, 5, 1]$ percentages.~\autoref{fig:metric_rankings} shows mean quality metric ranking over all datasets for each ensembling strategy, with lower values indicating better relative performance on that particular metric. We found that on average, $E_{\text{median}}$ performs comparably to individual UMAP runs on local, cluster-level, and global metrics (with the exception of neighborhood hit) indicating that ensembling does not significantly disrupt important properties from the original projections. We also found that adding 1\% and 5\% noise to input data reduced overfitting without greatly affecting quality metrics on average (as exemplified in~ \autoref{fig:problems_summary}). At higher levels of noise and with subsampling, however, ensembles performed worse across local quality metrics, but generally improved on several cluster-level and global measures. Notably, distance consistency was highest with 25\% added noise, and stress was generally lowest when ensembling projections of a 90\% sample rather than the full dataset. These findings suggest a tradeoff where perturbation may improve higher-level properties of ensembles (as in the case of reducing overfitting) at the cost of more fine-grained local relationships.

\subsection{Visualizing projection ensembles}
\label{sec:ensemble_vis}

We explored five methods for visualizing projection ensembles listed below, each with different interpretation goals.~\autoref{fig:ensemble_vis_types} shows examples of each method on the Human Activity Recognition dataset \cite{Anguita2013APD}.%, where each method surfaces different types of variability. 

\textbf{Median projection}: We plotted the median ensemble $E_{\text{median}}$ as a scatterplot. We encode average pointwise Procrustes distance with color to show which points move the most between runs. \textit{Applications}: this visualization helps avoid overinterpretation of individual NLDR runs, and the color encoding can help show \textit{how much} individual points are moving, but not their range of motion or higher-level cluster and structure changes across runs.

\textbf{Small multiples}: To select projections that best capture the range of possible outputs, we calculated the Procrustes distance from each generated projection to a reference and plot $n$ evenly spaced from the most to least distant. \textit{Applications}: In contrast to the median projection, small multiples allow for comparison of structure and clustering changes across runs (as in~ \autoref{fig:problems_summary}), but it is difficult to see movement in individual points across runs, particularly in larger datasets.

\textbf{Confidence ellipses:} For each point, we computed a confidence ellipse from their covariance over all projections. We plotted each mean (center) point and added ellipses corresponding to the first and third standard deviations by default, which contain approximately 39.4\% and 98.9\% of the positions for that point across all runs respectively. \textit{Applications}: As shown in~ \autoref{fig:ensemble_vis_types}, we found that confidence ellipses can help communicate uncertainty on multiple levels including the identification of outlier points with large confidence intervals, comparison of cluster variability based on the size and coherence of surrounding ellipses, and overall projection variability based on the average size of point covariances. For example, the two left-most clusters appear relatively stable based on their small surrounding ellipses, there are outlier points between the two clusters with high covariance, and the group of points on the right has greater overall movement, indicating higher uncertainty in the modeling process.

\textbf{Movement lines}: For all $1, \dots, i, i+1, \dots n$ projections, we plot lines between the position of each point in projections $i$ and $i+1$. We show the count of lines that cross each pixel position to allow scaling to larger datasets. We used a log-scale for the count value because we observed there are often areas of high-overlap which are not moving, and outlier areas with far fewer overlaps that are still potentially useful. \textit{Applications:}~ \autoref{fig:ensemble_vis_types} shows disorganized movement lines on the left side of the figure corresponding to outlier points that moved a large distance in almost every consecutive UMAP run. Unlike the median and confidence ellipses methods, we see \textit{how} these points were moving in addition to their range, but are not able to identifying individual non-outlier points or determine consensus positions for points. 
    
\textbf{KDE + binned points}: To aid in the identification of consistent clusters and densities, we plot points from all runs in a 2D histogram. Inspired by contour boxplots \cite{whitaker2013contour} we added contours around the smallest set of grid points that sum to 50\% and 95\% of the total density using a Gaussian KDE. We experimented with the KDE bandwidth on a set of 1000 MNIST UMAP projections and settled on the default (Silverman's rule of thumb \cite{silverman2018density}) for all experiments. \textit{Applications:} At the cost of showing individual points and point variability, the aim of this visualization is to highlight consistent trends in \textit{density} across runs including clusters and structure. We found it was useful to overlay both 2D bins and KDE contours in the case where contours are displayed around an area of low density. For example, in~\autoref{fig:ensemble_vis_types} there is a small contour in the top right of the figure that could correspond to a unique cluster, but upon closer inspection the 2D bins show there are very few points in that area.
\section{Discussion}
\label{sec:discussion}
In this work, we focus on two key interpretation problems with NLDR projections.  The first is that projection variance can produce substantial run-to-run differences, leading analysts to reach divergent interpretations depending on which projection they view. Exacerbating this problem, we found that the stability of NLDR projections can vary greatly over different datasets (\autoref{fig:procrustes_distances}). To reduce dependence on any single projection, we instead produce an ensemble of projections and visualize the median projection (Section~\ref{sec:ensemble_preprocessing}). We then visualize variation around this consensus to reveal which point structures are sensitive to stochastic variation (Section~\ref{sec:ensemble_vis} and ~\autoref{fig:ensemble_vis_types}). This exploratory work demonstrates the potential of ensemble-based approaches to reduce overinterpretation of individual projections and offers several candidate visualizations for communicating projection uncertainty. As described in Section~\ref{sec:ensemble_vis}, each ensemble visualization approach has distinct strengths and limitations. Future work is needed to evaluate which ensemble visualization methods most effectively support interpretation across different analytic tasks and contexts. 

The second interpretation problem we examined is that projections can appear highly stable even when they are overfitting noise. In these cases, consistency across runs may create false confidence that the visual structure reflects real patterns in the underlying data. To address this problem, we applied established perturbation techniques before generating the projection ensembles, using added noise and subsampling to test whether apparent structures remained stable under small changes to the input data (Section~\ref{sec:ensemble_preprocessing}). Our results suggest that modest perturbations can help reveal brittle and potentially spurious patterns without degrading the overall projection quality. Together, these approaches have the potential to communicate projection reliability and support more cautious interpretation of NLDR outputs. 

\bibliographystyle{abbrv-doi}
\bibliography{custom}

\pagebreak

\section*{Appendix}

\begin{figure}
    \centering
    \includegraphics[width=\linewidth,alt={A nearly identical heatmap to figure 4 in the main paper, showing little difference in quality metrics between mean and median ensemble projections.}]{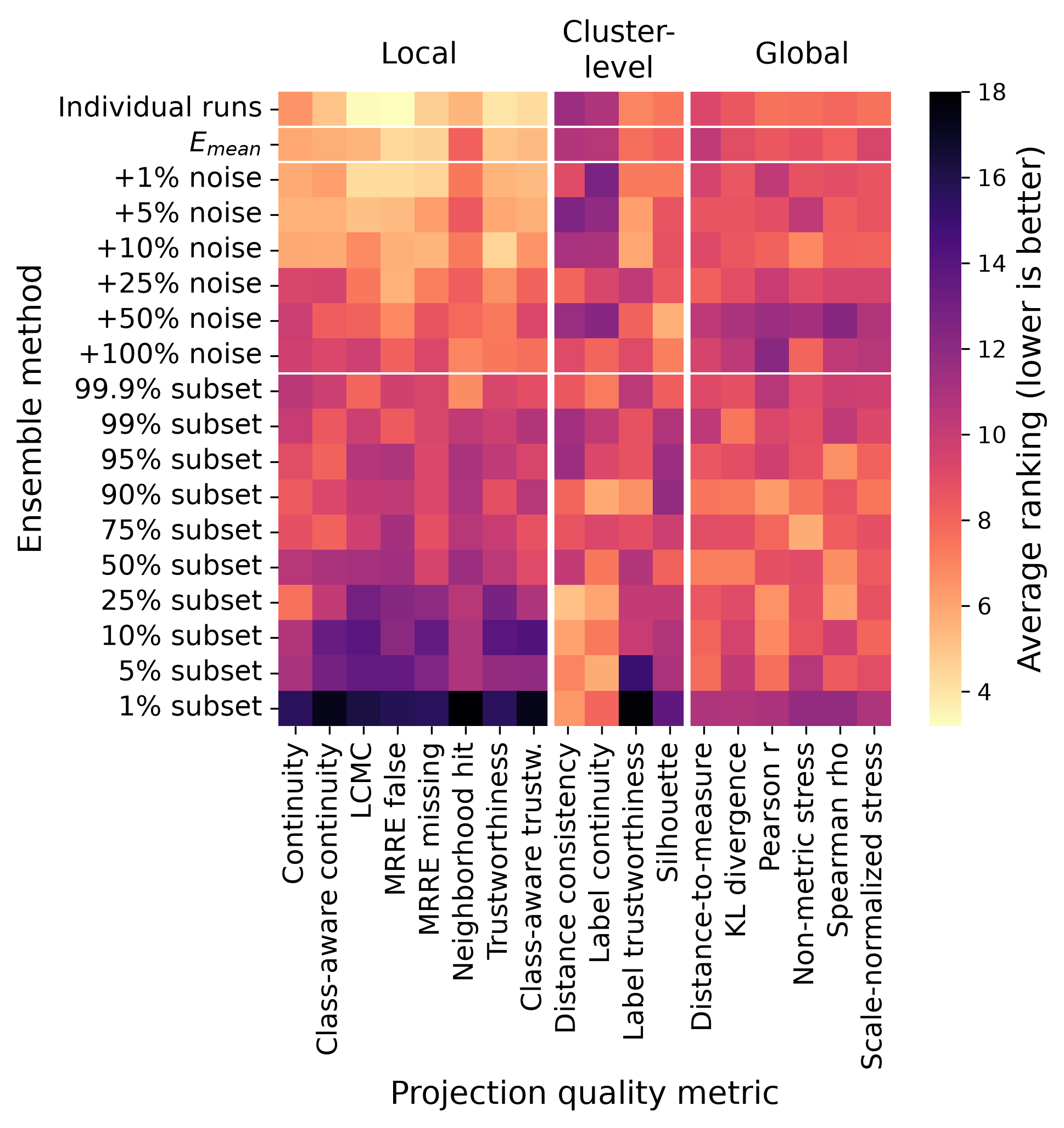}
    \caption{Average ensemble metric rankings for the mean ensemble ($E_\text{mean}$) compared to individual projections}
    \label{fig:emean_metric_rankings}
\end{figure}

Figure \autoref{fig:emean_metric_rankings} shows the average quality metric ranking of mean projection ensembles with different levels of noise and dropout, closely reproducing the results for $E_\text{median}$. \autoref{tab:datasets} below lists all 31 datasets used in our evaluation with citations. 

\begin{table*}[t!]
\centering
\begin{tabular}{|c|c|c|c|c|}
\hline
\rowcolor[HTML]{EFEFEF} 
\textbf{Dataset}                                  & \textbf{\# Samples} & \textbf{Dimensionality} & \textbf{\# Classes} & \textbf{\# Projections} \\ \hline
Circles \cite{scikit-learn}                     & 100                 & 2                       & 2                   & 1000                    \\
Moons \cite{scikit-learn}                       & 100                 & 2                       & 2                   & 1000                    \\
Iris \cite{iris_53}                            & 150                 & 4                       & 3                   & 1000                    \\
Wine \cite{wine_109}                           & 178                 & 13                      & 3                   & 1000                    \\
Circles 3d                     & 200                 & 3                       & 2                   & 1000                    \\
Olivetti faces \cite{Olivetti_faces}           & 400                 & 4096                    & 40                  & 1000                    \\
2d line                                              & 500                 & 2                       & N/A                 & 1000                    \\
Breast cancer \cite{breast_cancer_14}         & 569                 & 30                      & 2                   & 1000                    \\
Coil-20 \cite{nene1996columbia}                 & 1440                & 16384                   & 20                  & 1000                    \\
pbmc3k \cite{wolf2018scanpy}                    & 2700                & 32738                   & N/A                 & 100                     \\
s-curve with hole \cite{scikit-learn}           & 2860                & 3                       & N/A                 & 1000                    \\
Roll with hole \cite{scikit-learn}              & 3000          & 3                       & N/A                & 1000                    \\
Gaussian blob \cite{scikit-learn}                      & 3000                & 2                       & N/A                 & 1000                    \\
3 Gaussian blobs \cite{scikit-learn}                     & 3000                & 2                       & 3                   & 1000                    \\
%nd-blob \cite{scikit-learn}                     & 3000                & Various                 & N/A                 & 100                     \\
Samples from $U([0, 1]^3)$                                        & 3000                & 3                       & N/A                 & 1000                    \\
Samples from $U([0, 1]^4)$                                        & 3000                & 4                       & N/A                 & 1000                    \\
5k word vectors from Olmo2 \cite{olmo20242}             & 5013                & 2048                    & N/A                 & 1000                    \\
10\% of MNIST \cite{lecun2010mnist}            & 6000                & 784                     & 10                  & 1000                    \\
10\% of ambiguous MNIST \cite{mukhoti2021deterministic}           & 6000               & 784                     & $\sim$10            & 1000                    \\
HAR \cite{anguita2013public}                             & 7352                & 561                     & 6                   & 1000                    \\
11 100d spheres \cite{jeon2025umato}                 & 10000               & 101                     & 11                  & 1000                    \\
Mammoth \cite{smithsonian2020mammuthus}         & 50000               & 3                       & 10                  & 1000                    \\
MNIST \cite{lecun2010mnist}                    & 60000               & 784                     & 10                  & 1000                    \\
Fashion MNIST \cite{xiao2017fashion}            & 60000               & 784                     & 10                  & 1000                    \\
Ambiguous MNIST \cite{mukhoti2021deterministic}           & 60000               & 784                     & $\sim$10            & 1000                    \\
Samusik \cite{weber2016comparison}              & 86864               & 50                      & 25                  & 100                     \\
100k prime factorizations \cite{mcinnes2018umap}            & 100000              & 9592                    & N/A                 & 100                     \\
Levine \cite{weber2016comparison}               & 167044              & 13                      & 25                  & 100                     \\
Glove-6b \cite{pennington2014glove}             & 400000              & 300                     & N/A                 & 100                     \\
Allen ASAP pmdbs \cite{alleninstituteASAPHuman} & 2796736             & 100                     & 30                  & 10                      \\
HIGGS \cite{higgs_280}                         & 11000000            & 28                      & 2                   & 10                      \\ \hline
\end{tabular}
\caption{Datasets used in our evaluation of projection stability and quality metrics.}
\label{tab:datasets}
\end{table*}

\end{document}